\documentclass[a4paper,12pt]{article}

\usepackage{epsfig,graphicx,amsmath,amssymb}%,citesort}
\newcommand{\cO}{{\cal O}}
\newcommand*{\Mpp}{M_{\pi\pi}}
\begin{document}

\thispagestyle{empty}

$\phantom{.}$

\hfill

\begin{center}
{\Large {\bf PWA tools in Hadronic Spectroscopy } \\
\vspace{0.75cm}
{\huge {\bf  Mainz
}}}

\vspace{1cm}

{\large Feb 18--20, 2013 in Mainz, Germany}

\vspace{2cm}

\vspace{2.5cm}

ABSTRACT

\end{center}

\vspace{0.3cm}

\noindent
The mini-proceedings
of the Workshop on PWA tools in Hadronic Spectroscopy held in
Mainz from February 18$^{\rm th}$ to 20$^{\rm th}$, 2013.

\medskip\noindent
The web page of the conference, which contains all talks, can be found at
\begin{center}
{\tt http://conference.kph.uni-mainz.de/pwa2013/}
\end{center}

\vspace{0.5cm}

\noindent
We acknowledge the support of Deutsche Forschungsgemeinschaft DFG through the Collaborative Research Center ``The Low-Energy Frontier of the Standard Model" (SFB 1044).

\newpage

{$\phantom{=}$}

\vspace{0.5cm}

\tableofcontents

\newpage

\section{Introduction -- Scope of the workshop}

\addtocontents{toc}{\hspace{1cm}{\sl M.~Ostrick, M.~Fritsch, and L.~Tiator}\par}

\vspace{5mm}

\noindent
M.~Ostrick, M.~Fritsch, and L.~Tiator

\vspace{5mm}

\noindent
Institut f\"ur Kernphysik, Johannes Gutenberg-Universt\"at Mainz, Germany\\

\subsection{Motivation}

The understanding of scattering amplitudes and
the interpretation of resonances is one of the most challenging
tasks in modern hadron physics. Usually, these resonances are interpreted
in terms of excitation spectra of hadrons. In this case
the properties of a resonance, defined by a pole of the scattering amplitude
in the complex energy plane, are related to universal properties of a hadron.
The position of these poles are independent of the initial and final
state of a particular reaction and should be compared to the spectrum predicted
by QCD. In order to determine such resonance parameters in different reactions,
an immense experimental effort started during the last decade.

There is an unprecedented increase in high precision measurements of
single and double spin observables in photo-induced meson production
at ELSA, JLAB and MAMI. Even if the experiments are still ongoing it
is already obvious from preliminary results that these data will
have a significant impact on our knowledge of baryon resonances.

Completely different approaches to hadron spectroscopy are presently
being developed at the BES-III $e^+e^-$ collider , where decays like
$J/\psi \to \bar N N^* \to \bar N N \pi$  have been observed, or at
the COMPASS experiment at CERN, where high statistics measurements
of diffractive $pp$, $\pi p$ and $K p$ collisions clearly show
baryon and meson resonances in different final states. Furthermore,
the HADES experiment at GSI started to study resonances in direct
$pp$ collisions.

One important milestone in hadron spectroscopy will be the combination
of all this new empirical information from very different experiments.
This requires the mutual understanding of the techniques and model assumptions
used by the different communities in data analysis and for the extraction
of partial wave amplitudes and resonance parameters.
The aim of this workshop was to deepen this mutual understanding and to start a discussion
about the development of common tools and analysis techniques.
\newpage

\section{Short summaries of the talks}

\subsection{Improved Breit-Wigner Formula}
\addtocontents{toc}{\hspace{2cm}{\sl S.~Ceci}\par}

\vspace{5mm}

\noindent
S.\ Ceci, M.\ Korolija, and B.\ Zauner

\vspace{5mm}

\noindent
Institut Rudjer Bo\v skovi\' c, Bijeni\v cka 54, HR-10000 Zagreb, Croatia\\

\vspace{5mm}

Fundamental properties of a simple narrow resonance are the mass $M$, roughly equal to the peak position of the cross section, and the decay width $\Gamma$, which is related to the width of the bell shape. These parameters, along with the branching fraction $x$, are known as the Breit-Wigner parameters \cite{BW}.  However, resonance peaks are usually very wide, and the shape can be so deformed that it is not at all clear where exactly is the mass, or what would be the width of that resonance. In those cases, resonance parameters are assumed to have energy dependence. This dependence is often defined differently for different resonances.

With model dependent parameterizations, the simple connection between
physical properties of a resonance and its model parameters is lost, and the
choice of the ``proper" resonance parameters becomes the matter of
preference. This makes the comparison
between cited resonance parameters quite confusing and potentially hinders
the direct comparison between microscopic theoretical predictions (for example~\cite{Dur08}), and experimentally obtained resonance properties \cite{PDG}. To clarify this situation we devised a simple model-independent formula for
resonant scattering, with well defined resonance physical properties, which
will be capable of successfully fitting the realistic data for broad
resonances \cite{Cec13}.

These Breit-Wigner parameters are uniquely defined and model independent,
with directly observable mass as the peak of the squared amplitude (a $T$-matrix element or partial wave).
However, they strongly depend on shape parameter that corresponds to the complex residue phase, and which may change from reaction to reaction. Consequently, for the same pole position, there
will be different Breit-Wigner masses and widths in different channels.

In Ref.~\cite{Cec13}, we have shown that the original Breit-Wigner formula can be
considerably improved by including a single additional phase parameter.  Moreover, our results suggest that this parameter seems to be equal to the half of the resonance residue phase, regardless of the
resonance inelasticity.  The improved Breit-Wigner formula has two equivalent forms that can
be used to estimate either the pole, or the Breit-Wigner parameters in a
model independent way. Analysis of the results for the both sets of the resonance parameters has shown us that in the  PDG tables \cite{PDG} there are values that do not correspond either to
pole, or to Breit-Wigner values.  Such an outcome undermines the proper
matching between microscopic theories (such as the lattice QCD \cite{Dur08}) and
experiment, and should be properly addressed in the PDG notes.

\newpage

\subsection{Coupled-channel dynamics in meson-baryon scattering}
\addtocontents{toc}{\hspace{2cm}{\sl M.~D\"oring}\par}

\vspace{5mm}

\noindent
M. D\"oring

\vspace{5mm}

\noindent

Helmholtz-Institut f\"ur Strahlen- und Kernphysik (Theorie) and Bethe Center for Theoretical
Physics,  Universit\"at Bonn, Nu\ss allee 14-16, D-53115 Bonn, Germany\\

\vspace{5mm}

    Mapping out the spectrum of excited baryons from data is a cornerstone to the understanding of the strong interaction. With the advent of high-precision photo and electroproduction data [1-3] there is finally the chance to come to more conclusive answers regarding the resonance content, that have been searched for since a long time [4-18]. Also, the dynamical nature of several baryonic resonances remains subject of investigation [19-22].

    One challenge in the study of excited baryons is to combine the data of different reactions in a global analysis while respecting unitarity, analyticity [23,24], gauge invariance [25,26] and other theoretical requirements such as chiral constraints. For example, if branch points in the complex plane are missing in an analysis tool, their absence might be masked by erroneous resonance signals [27].

    Recent results of the analysis of pion photoproduction [25] and the world data of pion-induced $\pi N$, $\eta N$, $K\Lambda$, and $K\Sigma$ production are presented [28], developed at the Forschuns\-zentrum J\"ulich, the Universities of Bonn and of Athens (GA), and The George-Washington University [23-29]. The amplitude allows for the reliable extraction of resonance poles and residues that are determined up to $J^P=9/2^\pm$.

    Universal principles like (coupled-channel) unitarity also apply to the analysis of upcoming data of resonances decaying in the finite volume of lattice simulations. As quark masses drop towards physical values the eigenvalue spectrum becomes dominated by finite-volume effects. Hadronic methods such as coupled-channel analysis [30] can be adapted to the analysis of eigenvalues, in particular when moving frames lead to additional partial wave mixing [31], or if chiral extrapolations are required [32]. Effective Lagrangian methods [33] or unitarized chiral interactions [34] can provide adequate frameworks.

\newpage

\subsection{Using the Bonn-Gatchina PWA to constrain the production of an $\bar{K}$NN-cluster in p+p collisions}
\addtocontents{toc}{\hspace{2cm}{\sl E.~Epple}\par}

\vspace{5mm}

\noindent
E. Epple for the HADES collaboration

\vspace{5mm}

\noindent
Excellence Cluster ``Universe", TU M\"unchen, Garching Germany\\

\vspace{5mm}
In the given talk, I focused on the reaction:
\begin{equation}
p+p \rightarrow p + K^{+} + \Lambda, \label{rea:1}
\end{equation}
at a beam kinetic energy of 3.5 GeV. %(Measured by HADES usw.)
This reaction is the simplest way for open strangeness production in p+p collisions.
Our particular aim is to study the possible formation of an intermediate ``$ppK^{-}$" bound-state and its coupling to the $p+\Lambda$ final state:
%It is of special interest for us, as it may contain the end-products of the intermediate
%production of a kaonic cluster:
\begin{equation}
p+p \rightarrow \bar{K}NN + K^{+}\rightarrow p  + \Lambda + K^{+}.
\end{equation}
The possible existence of bound states of $\bar{K}$ and nucleons was first mentioned in \cite{Wycech:1986yb}
but this issue triggered theoretical efforts only when Y. Akaishi and T. Yamazaki predicted that due to the strong binding between the kaon and the nucleons extreme dense systems might be formed \cite{Akaishi:2001zb,Akaishi:2002bg}.
Meanwhile, several theoretical investigations have been performed by various groups, confirming the existence of such states
but giving also a wide range of predictions for possible masses and width of the simplest bound state of a kaon and two nucleons ($\bar{K}NN$), see \cite{Gal:2013vx} and references therein.
\par
The question, whether such a new state is produced in Reaction (\ref{rea:1}) or not can only be answered if one understands the standard production process of $p+K^{+}+\Lambda$ in proton proton collisions.
Here, several experiments have shown that the final state of $K^{+}+\Lambda$ is dominated by the decay of $N^{*+}$ resonances \cite{Bierman1966,Chinowsky1968,Cleland1984,Schulte-Wissermann,AbdEl-Samad2010}.
This fact heavily influences the dynamics of the reaction. In order to find an additional signal beyond the known $p+K^{+}+\Lambda$ production we have to determine the properties of the $N^{*+}$-resonances that are the main source of the $K^{+}+\Lambda$ yield.
\par
\begin{figure}[t]
\begin{center}
\begin{minipage}[t]{0.48\textwidth}
\centerline{\epsfysize 5.9 cm
\epsfbox{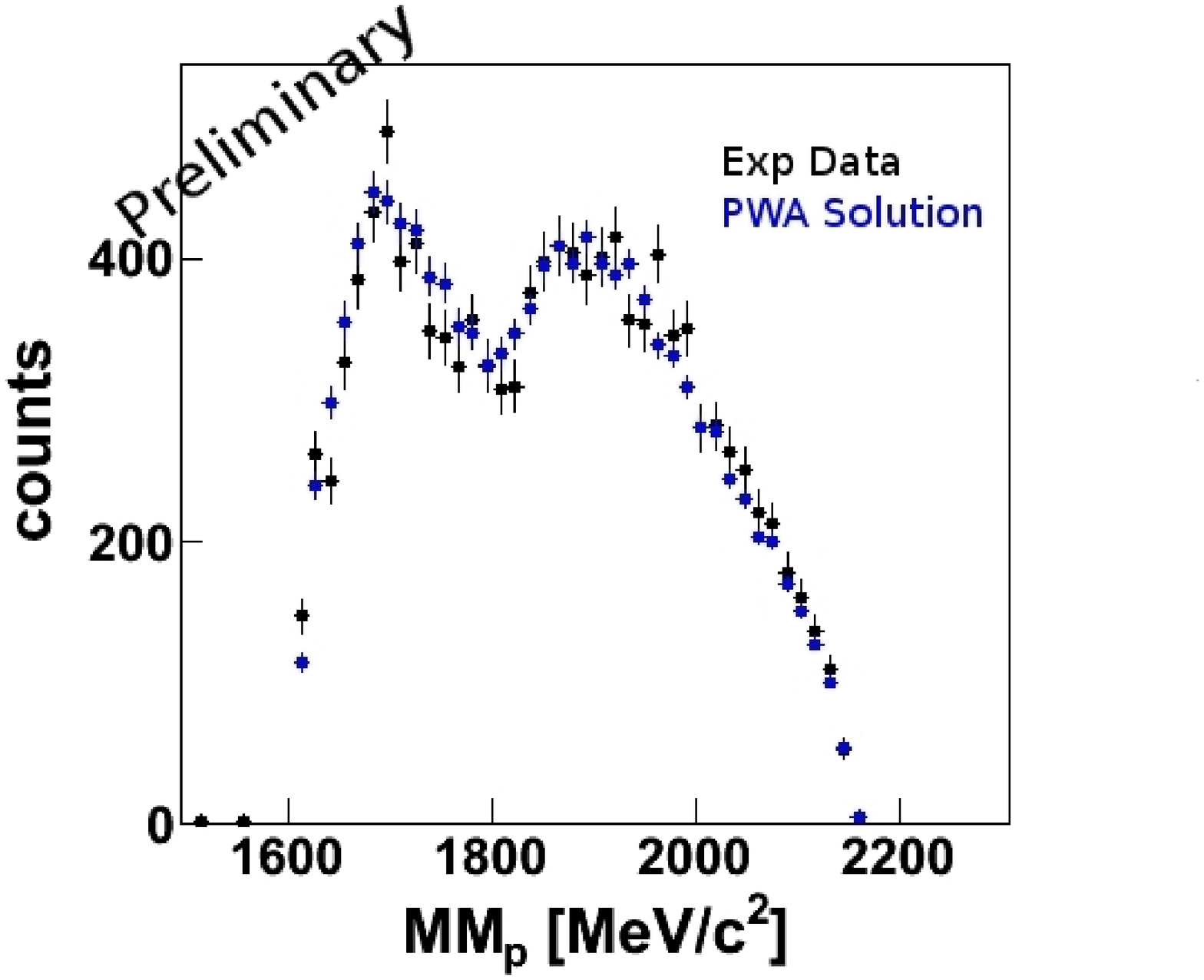}}
\caption{\label{Fig:1}$\Lambda K^{+}$ invariant mass. Data with a PWA solution.}
\end{minipage} \hfill
\begin{minipage}[t]{0.48\textwidth}
\centerline{\epsfysize 5.9 cm
\epsfbox{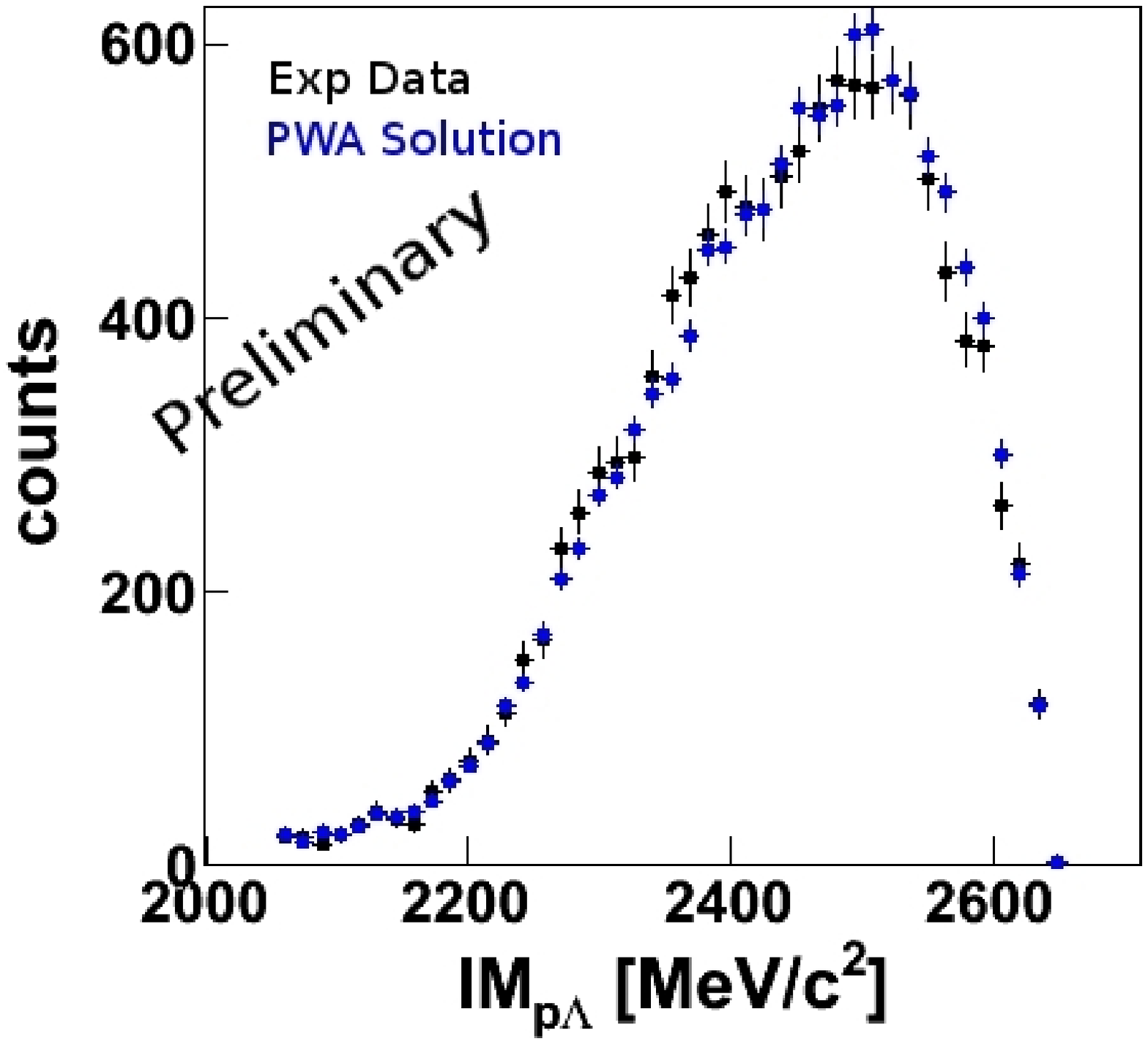}}
\caption{\label{Fig:2} Invariant mass of $\Lambda$p compared to a PWA solution.}
\end{minipage}
\end{center}
\end{figure}

To cope with this task, we have used the Bonn-Gatchina Partial wave analysis framework \cite{a3}.
Besides ordinary non-resonant production of $p+K^{+}+\Lambda$ the following final state is included in the framework:
\begin{equation}
p+p \rightarrow p + N^{*+}\rightarrow p  + \Lambda + K^{+}.
\end{equation}
We have scanned the PDG tables \cite{Beringer:1900zz} to find suited candidates for $N^{*+}$
resonances.
The following resonances were selected: $N(1650)$, $N(1710)$, $N(1720)$, $N(1875)$, $N(1880)$,
$N(1895)$ and $N(1900)$ as the energy of 3.5 GeV is above their production threshold in proton+proton collisions and as they have a measured decay branching into $K^{+}+\Lambda$.
The data that we interpret with help of the PWA were collected with the HADES spectrometer
\cite{Agakishiev:2009am}. Out of 1.2 $\cdot 10^{9}$ p+p collisions two data-samples corresponding to the final state $p+K^{+}+\Lambda$, with a statistic of 13,000 and 8,000 events, could be
extracted.
The 21,000 events were fitted with the sum of several transition amplitudes including resonant
and non-resonant production of the tree final state particles. To account for the fact that there
is an ambiguity which resonances participate in the process, a systematic variation of the number
of included resonant and non-resonant waves was performed.
Figures \ref{Fig:1} and \ref{Fig:2} show the best PWA solution for the data set with 13,000
events. The solution can describe the data remarkably well.
\par
The PWA solution is a model for the background-only hypothesis of the analysis, as it includes no production of a kaonic cluster.
One can perform a statistical test of this hypothesis to find incompatibilities of the hypothesis with the data which might be
explained with the the presence of an additional signal in the data \cite{Beaujean:2011zza}.
Thus, we have computed a local p-value as a function of the $p\Lambda$ invariant mass, which has revealed that
the data are compatible with the background-only hypothesis.
We will, thus, concentrate on extracting an upper limit of the production cross section of a kaonic cluster
produced in p+p collisions at 3.5 GeV with a branching in $p\Lambda$.
As the expected cross section is very small we intend to use the CL$_{s}$ method for this purpose \cite{Read:2002hq}.

\newpage

\subsection{Partial Wave Extraction from Near-Threshold Neutral Pion Photoproduction Data}
\addtocontents{toc}{\hspace{2cm}{\sl C. Fern\'andez-Ram\'{\i}rez}\par}

\vspace{5mm}

\noindent
C. Fern\'andez-Ram\'{\i}rez

\vspace{5mm}

\noindent
Grupo de F{\'i}sica Nuclear, Universidad Complutense de Madrid, CEI Moncloa, Spain\\

\vspace{5mm}
Since the development of chiral perturbation theory in the nucleon sector \cite{Gasser:1987rb,Bernard:1995dp},
neutral pion photoproduction in the near-threshold region \cite{Bernstein:1996vf,Schmidt:2001vg}
has played a pivotal role in our understanding of chiral symmetry
\cite{Bernard:1992nc,Bernard:1994gm,Bernard:2001gz,FernandezRamirez:2009su,FernandezRamirez:2009jb,Fuhrer:2009gj}
and, henceforth, Quantum Chromodynamics in the low energy regime \cite{Book}.

The latest experiment by the A2 and CB-TAPS collaborations at MAMI/Mainz
has measured the energy dependence of the photon beam asymmetry together
with the differential cross section of the $\vec{\gamma} p \to \pi^0 p$ process
allowing to extract the relevant partial waves in an almost model independent way \cite{Hornidge:2012ca}.
With these high quality data it has been possible to assess the energy range
where state-of-the-art Heavy Baryon Chiral Perturbation Theory \cite{FernandezRamirez:2012nw}
and Relativistic Chiral Perturbation Theory \cite{Hilt:2013uf} can be applied,
obtaining an upper limit of $\sim 170$ MeV of photon energy in the laboratory frame.

The next step to improve the agreement between theory and experiment in the near threshold region
is the explicit inclusion of the $\Delta$(1232) resonance in the chiral
calculations \cite{Hemmert:1997ye,Hemmert:1996xg,Lensky:2009uv,Alarcon:2011zs,Alarcon:2012kn}.

Experimental data on the F and T asymmetries are currently under analysis \cite{Schumann:2012zz}
and will, hopefully, allow to test our knowledge on D waves as well as to obtain the $\beta$ parameter
associated to unitarity \cite{Bernstein:1998ip}.

\newpage

\subsection{Truncated PWA of a complete experiment for photoproduction of two
pseudoscalar mesons}
\addtocontents{toc}{\hspace{2cm}{\sl A.~Fix}\par}

\vspace{5mm}

A.~Fix$^1$ and H.~Arenh\"ovel$^2$

\vspace{5mm}

\noindent
$^1$Tomsk Polytechnic University, Russia\\
$^2$Institut f\"ur Kernphysik, Johannes Gutenberg-Universt\"at Mainz,
Germany\\

\vspace{5mm}

A truncated partial wave analysis for the photoproduction of two pseudoscalar
mesons on
nucleons is discussed. For the partial wave decomposition of the amplitude the
method developed in Ref.~\cite{FiA12} is used. Here a special
coordinate frame is chosen in which the $z$-axis is taken along the normal to
 the final state plane
spanned by the momenta of the final particles. In this case one can take
as independent
kinematical variables the energies of two of the three final particles
and three
angles, determining the orientation of the final state plane with respect
to the photon
beam. Then the amplitudes corresponding to a definite value of the total
angular momentum $J$ and its
projection $M$ on the $z$-axis are obtained through an expansion of the
helicity
amplitudes over a set of Wigner functions (rotation matrices).

To find a complete set of observables, we have applied a criterion,
which has been developed previously
for photo- and
electrodisintegration of a deuteron \cite{ArLeiTom,ArLeiTomFBS}.
It allows one to find a 'minimal' set in order to
determine the partial wave amplitudes up to possible discrete ambiguities.
To resolve the remaining ambiguities a second criterion from
Ref.\,\cite{Tabakin} is
adopted for the truncated PWA. In the simplest case of $J_{max}=1/2$ these
two criteria turn out to be sufficient
for an unambiguous determination of the eight independent partial wave
amplitudes. The
corresponding 'fully' complete set includes beyond the unpolarized cross
section,
helicity beam asymmetry, as well as nucleon recoil polarization along the
$z$ and one of the
$x$ or $y$ axes with and without circular polarization of the photon beam.

\newpage

\subsection{Common PWA Framework for PANDA}
\addtocontents{toc}{\hspace{2cm}{\sl M.~Fritsch}\par}

\vspace{5mm}

Miriam Fritsch on behalf of the ComPWA group

\vspace{5mm}

\noindent
Helmholtz Institut Mainz und Institut f\"ur Kernphysik, 
Johannes Gutenberg-Universt\"at Mainz, Germany\\

\vspace{5mm}

The spectroscopy program of future high-precision experiments like PANDA at FAIR 
implies the study of creation processes and the analysis of the distribution of 
hadronic decay particles to understand the nature of hadrons, as one cannot 
study their quark content directly due to confinement. Reaching unprecedented 
statistics and excellent resolution with new experiments a 
detailed and complex modeling of the strong interaction processes is required. 
To determine the quantum numbers of resonances and suppress non-resonant 
background as well as contributions from other waves, one has to deconvolute 
the final state of interest into separate spin-parity components by performing
a partial wave analysis (PWA). As the number of parameters to be 
determined by the fit 
increases rapidly
with the number of possible waves and therefore with the energy of the 
system, PANDA most likely will have to deal with hundreds of free parameters 
per fit. This makes the analyses a very demanding exercise in terms of 
computing resources, analysis management and quality assurance.

Currently available software is often tailored to the needs 
of one experiment or even specific channels requiring a large effort to 
use it for other needs e.g. other initial states or specific formalisms.
Therefore, a new software package should also consider the latest and future 
developments in theory and provide a modular framework to make use of 
new or modified modeling approaches.

The Common PWA Framework for PANDA is designed to avoid the limitations 
of the available
software packages and to take into account the experience from the whole 
community. Therefore we conducted a comprehensive survey and collected 
the requirements 
for this software package in a document which specifies the capabilities of
the framework and defines the 
different modules. The main requirements were: Separation of physics models, 
data handling and fitting techniques, usage in multiple experiments, 
simultaneous fitting of multiple data sets,
extendable with various models and formalisms,
full freedom of modeling of physics amplitudes,
easy usage of various optimization routines and
caching and parallelization in different stages.
To provide a playground for the framework design, a first environment with 
a simple fit routine was set up. It will be used as a base for further 
extensions, provides basic tools and most importantly it demonstrates the 
modular design.

\newpage

\subsection{Experimental Studies of the N* Structure with CLAS and CLAS12}
\addtocontents{toc}{\hspace{2cm}{\sl R.W.~Gothe}\par}

\vspace{5mm}

\noindent
R.W.~Gothe

\vspace{5mm}
\noindent
Department of Physics and Astronomy, University of South Carolina, Columbia, USA\\

\vspace{5mm}
Baryon spectroscopy can establish more sensitively, and in an almost
model-independent way, nucleon excitations and non-resonant reaction
amplitudes by complete measurements of pseudo-scalar meson photoproduction
off nucleons. However, beyond baryon spectroscopy at the real photon point
$Q^2=0\,(GeV/c)^2$, electron scattering experiments can also investigate
the internal hadronic structure at various distance scales by tuning
the four-momentum transfer from $Q^2 \approx 0\,(GeV/c)^2$, where the
meson cloud contributes significantly to the baryon structure, over
intermediate $Q^2$, where the three constituent-quark core starts to
dominate, to $Q^2$ up to $12\,(GeV/c)^2$, attainable after the
$12\,GeV$ upgrade at JLab, where the constituent quark
gets more and more undressed towards the bare current quark
\cite{Az12,Bh03}. Although originally derived in the high $Q^2$
limit, constituent counting rules describe in more general terms how
the transition form factors and the corresponding helicity amplitudes
scale with $Q^2$ dependent on the number of effective constituents.
Recent results for $Q^2 < 5\,(GeV/c)^2$ \cite{Mo12,AzB09,Par08}
already indicate for some helicity amplitudes, like
$A_{1/2}$ for the electroexcitation of the $N(1520)D_{13}$,
the onset of proper scaling assuming three constituent quarks. This
further indicates that in this case the meson-baryon contributions
become negligible in comparison to those of a three constituent-quark
core, which coincides nicely with the EBAC dynamical coupled channel
calculation \cite{Az12,Ma07,Sa96}. Along the same line of reasoning
perturbative QCD (pQCD) predicts in the high $Q^2$-limit by neglecting
higher twist contributions that helicity is conserved. The fact that
this predicted behavior sets in at much lower $Q^2$ values than
expected \cite{Az12,AzB09} challenges our current
understanding of baryons even further. For $N(1520)D_{13}$ the
helicity conserving amplitude $A_{1/2}$ starts to dominate the
helicity non-conserving amplitude $A_{3/2}$ at $Q^2 \approx
0.7\,(GeV/c)^2$, as typically documented by the zero crossing
of the corresponding helicity asymmetry
$ A_{hel} = (A_{1/2}^2 - A_{3/2}^2)/(A_{1/2}^2 + A_{3/2}^2)$.
The $N(1685)F_{15}$ resonance shows a similar behavior with a zero
crossing at $Q^2 \approx 1.1\,(GeV/c)^2$, whereas the
$\Delta(1232)P_{33}$ helicity asymmetry stays negative with no
indication of an upcoming zero crossing; and even more surprising are
the preliminary results for the $N(1720)P_{13}$ $A_{1/2}$ amplitude,
which decreases so rapidly with $Q^2$ that the helicity asymmetry
shows an inverted behavior with a zero crossing from positive to
negative around $Q^2 \approx 0.7\,(GeV/c)^2$ \cite{Mok10}.
These essentially different behaviors of resonances underlines that it
is necessary but not sufficient to extend the measurements of only the
elastic form factors to higher momentum transfer. Hence
to comprehend QCD at intermediate distance scales where dressed quarks
are the dominating degrees of freedom and to explore interactions of
dressed quarks as they form various baryons in distinctively different
quantum states, the $Q^2$ evolution of exclusive transition form
factors to various resonances up to $12\,(GeV/c)^2$ are absolutely
crucial \cite{GM09}.

The status quo in the baryonic structure analysis endeavor are
currently based on three corner stones. First, the analysis of the $N\pi$
channel data is carried out in two phenomenologically different approaches
based on fixed-$t$ dispersion relations and a unitary isobar
model~\cite{AzB09,Az03}. The main difference between the two
approaches is the way the non-resonant contributions are derived.
The $p\pi^+\pi^-$  CLAS data is analyzed within a unitarized
phenomenological meson-baryon
model \cite{Mo12,MoB09} that fits nine independent differential
cross sections of invariant masses and angular distributions. The good
agreement of the resonant helicity amplitude results in the single-
and double-pion channels, that have fundamentally different non-resonant
contributions and model-dependencies, provides evidence for the reliable
extraction of the $\gamma_vNN^*$ electrocoupling amplitudes.
Second, the high $Q^2$ behavior is most consistently described by
relativistic light-front quark models, like
\cite{Az12,Az07,Gia08,Cap95,Web90}, but their description of the
low $Q^2$ behavior is less satisfactory; and new QCD-based approaches,
such as Lattice QCD and Dyson-Schwinger equations of QCD, start to
capture the dynamical generation of mass \cite{Az12}. Third, the Excited
Baryon Analysis Center (EBAC), now the ANL-Osaka collaboration,
calculates, based on a full dynamical coupled-channel analysis
\cite{Az12,Ma07,Sa96}, meson-baryon dressing (meson-cloud)
contributions that seem to bridge the gap between the relativistic
light-front quark models and the measured results at low $Q^2$.

Digging deeper into the baryonic structure by increasing the momentum
transfer beyond $5\,(GeV/c)^2$ \cite{GM09} opens a unique window to
investigate the dynamic momentum-dependent structure of the constituent
quarks, and will allow us to address the central and most challenging
questions in present-day hadron physics: a) how more than 98\% of hadron
masses are generated non-perturbatively, b) how quark/gluon confinement
and dynamical chiral symmetry breaking emerge from QCD, and c) how the
non-perturbative strong interaction generates the ground and excited
nucleon states with various quantum numbers from quarks and gluons.

\newpage

\subsection{Partial-Wave Analyses at COMPASS}
\addtocontents{toc}{\hspace{2cm}{\sl B.~Grube}\par}

\vspace{5mm}

\noindent
B. Grube for the COMPASS Collaboration

\vspace{5mm}
\noindent
Physik-Department E18, Technische Universit\"at M\"unchen, Germany \\

\vspace{5mm}
(COMPASS)~\cite{compass} is a multi-purpose fixed-target experiment at
the CERN Super Proton Synchrotron (SPS) aimed at studying the
structure and spectrum of hadrons. One main goal of COMPASS is to
search for mesonic states beyond the constituent quark model that are
predicted by models and lattice QCD~\cite{beyondCqm}.
%Particularly
%interesting are so-called spin-exotic mesons which have $J^{PC}$
%quantum numbers forbidden for ordinary $q\bar{q}$ states.
During 2008 and 2009 COMPASS acquired large data samples of
diffractive-dissociation and central-production reactions using
190~GeV pion and proton beams on hydrogen and nuclear targets. The
focus of the COMPASS analyses lies on pion-induced single-diffractive
reactions of the form $\pi^- + \mbox{A} \to X^- + \mbox{A}$, where the
intermediate state $X^-$ decays into the final states $\pi^-\eta$,
$\pi^-\eta'$, $\pi^-\pi^+\pi^-$, $\pi^-\pi^0\pi^0$, $\pi^-\eta\eta$,
or $\pi^-\pi^+\pi^-\pi^+\pi^-$. At COMPASS energies these reactions
are dominated by Pomeron exchange. In the partial-wave analyses (PWA)
it is assumed that production and decay of the $X^-$ factorize and
that they can be described by two amplitudes $T_i$ and $A_i$,
respectively. Using the isobar model~\cite{isobar} and the helicity
formalism~\cite{helicity} the decay amplitudes $A_i$ can be calculated
with no free parameters.
%In the isobar model multi-body decays are
%described as a chain of sequential two-body decays.
The intermediate resonances that appear in the isobar model are
usually described using relativistic Breit-Wigner parametrizations
that include Blatt-Weisskopf centrifugal barrier penetration
factors~\cite{qh}. For intermediate $\pi\pi_{S\mbox{\scriptsize
    -wave}}$ isobars the $M$ solution from~\cite{pipi} is used. This
parameterization does not include the $f_0(980)$ which is added as a
separate independent isobar. Parity conservation at the $X^-$
production vertex is taken into account by using the reflectivity
basis~\cite{refl}. In the high-energy limit the reflectivity quantum
number corresponds to the naturality of the exchange particle in the
used Gottfried-Jackson $X^-$ rest frame~\cite{naturality}.

The mass dependence of the production amplitudes $T_i(m_X)$ is
determined by extended maximum likelihood fits to the observed
multi-dimensional kinematic distributions of the final state
particles. These fits are performed in narrow bins of $m_X$ taking
into account interference effects and detector acceptance. In the case
of the two-pseudoscalar final states, mathematical ambiguities arise
that have to be taken into account~\cite{ambi}. In a second step the
mass dependence of the spin-density matrix, which is given by the
production amplitudes determined in the first fit, is fitted by a
coherent sum of Breit-Wigner and background terms.

Two PWA packages are used for the analyses: One is a Fortran program
that was originally developed at Illinois~\cite{illinois} and later
modified at Protvino and Munich. The other one~\cite{rootpwa} is a C++
framework originally based on the PWA2000 package~\cite{pwa2000}.

First results show significant intensity in waves with spin-exotic
$J^{PC} = 1^{-+}$ quantum numbers, which are forbidden for ordinary
$q\bar{q}$ states, in the $\pi^-\eta$ and $\pi^-\eta'$~\cite{etaPi} as
well as in the $\pi^-\pi^+\pi^-$ and $\pi^-\pi^0\pi^0$ decay
channels~\cite{3pi}. Their resonance interpretation is, however, still
unclear. The data seem to contain significant background contribution
from non-resonant Deck-like processes~\cite{deck} that need to be
understood.

\newpage

\subsection{Theory of Two-Pion Production}
\addtocontents{toc}{\hspace{2cm}{\sl H.~Haberzettl}\par}

\vspace{5mm}

\noindent
H.~Haberzettl

\vspace{5mm}

\noindent
Institute for Nuclear Studies and Department of Physics.\\
The George Washington University, Washington, DC 20052, U.S.A.\\

\vspace{5mm}

A field-theory description of the photoproduction of two pions off the nucleon
is presented that applies to real as well as virtual photons in the one-photon
approximation. The approach is an extension of the field theory for $\gamma
N\to \pi N$ of Haberzettl~\cite{Haberzettl1997} whose practical
implementation~\cite{HNK2006,HHN2011} was recently shown~\cite{Huang2011} to
provide an excellent description of observables over a wide range of energies.

The Lorentz-covariant theory for $\gamma N\to \pi\pi N$ is complete at the
level of all explicit three-body mechanisms of the interacting $\pi\pi N$
system based on three-hadron vertices. The modifications necessary for
incorporating $n$-meson vertices for $n\ge 4$ are discussed. The resulting
reaction scenario subsumes and surpasses all existing approaches to the problem
based on hadronic degrees of freedom. The full three-body dynamics of the
interacting $\pi\pi N$ system is accounted for by the Faddeev-type ordering
structure of the Alt-Grassberger-Sandhas equations~\cite{AGS67}. The
formulation is valid for hadronic two-point and three-point functions dressed
by arbitrary internal mechanisms
--- even those of the self-consistent nonlinear Dyson-Schwinger type (subject
to the three-body truncation) --- provided all associated electromagnetic
currents are constructed to satisfy their respective (generalized)
Ward-Takahashi identities~\cite{Haberzettl1997,HNK2006,HHN2011}. The latter is
a \emph{necessary} condition for maintaining microscopic consistency of all
contributing electromagnetic reaction mechanisms. Following the basic
prescription given in Ref.~\cite{Haberzettl1997}, it was shown in
Refs.~\cite{HNK2006,HHN2011}, in particular, how the necessary full off-shell
behavior of the Ward-Takahashi identities can be restored even if, for
practical reasons, some of the underlying reaction dynamics needs to be
truncated because of its complexity.

It is shown that coupling the photon to the Faddeev structure of the underlying
hadronic two-pion production mechanisms results in a natural expansion of the
full two-pion photoproduction current $\Mpp^\mu$ in terms of multiple loops
involving two-body subsystem scattering amplitudes of the $\pi\pi N$ system
that preserves gauge invariance as a matter of course order by order in the
number of loops. A closed-form expression is presented for the entire
gauge-invariant current $\Mpp^\mu$ with complete three-body dynamics.
Individually gauge-invariant truncations of the full dynamics most relevant for
practical applications at the no-loop, one-loop, and two-loop levels are
discussed in detail.

The full description of the theory is forthcoming~\cite{HNO2pi}.

  \textit{Acknowledgments:} This work was supported in part by the National Research Foundation of Korea
  funded by the Korean Government (Grant No.~NRF-2011-220-C00011).

\newpage

\subsection{Remarks about resonances}
\addtocontents{toc}{\hspace{2cm}{\sl C.~Hanhart}\par}

\vspace{5mm}

\noindent
 C. Hanhart

\vspace{5mm}

\noindent
Institut f\"{u}r Kernphysik, J\"ulich Center for Hadron Physics and
 Institute for Advanced Simulation,  Forschungszentrum J\"{u}lich,  D-52425 J\"{u}lich, Germany

\vspace{5mm}
The presentation was a personal collection of general remarks about properties
of resonances and what it takes to extract them reliably from data.
Most illustrating examples were taken from the meson sector.

The central statement is that resonances are uniquely defined via their
pole positions and residues. Based on residues it is, e.g., possible to
define a branching ratio even for broad resonances by using the
narrow width formula also in this case --- this method was proposed
to quantify $\sigma\to \gamma\gamma$ in Ref.~\cite{pennington}
and is also proposed for baryons in Ref.~\cite{pdg}, p. 1269.
However, one should not be tempted to use pole parameters and residues
directly in the parametrization of the $T$--matrix, if a parametrization of
the amplitude over a larger energy range is the goal, for such a parametrization
of Breit-Wigner type violates unitarity and analyticity --- e.g. it will automatically
produce complex scattering lengths.

A better method is to use a parametrization that is consistent with unitarity
by construction as done in Ref.~\cite{caprini}. In this work it is shown that
even the pole of the $\sigma$ gets quite well constrained, if one starts
from theoretically well motivated parametrizations --- that include the left
hand cut as well as the two lowest right hand cuts --- and fits high quality data,
available in a relatively small energy range. The resulting scatter in pole positions
is comparable to that of phenomenological studies that also describe the low energy
data --- c.f. Ref.~\cite{pdg}, p. 708. In case of $\pi\pi$--scattering it is possible to even further improve the
quality of the pole determination by using Roy equations~\cite{roy}.

For a proper extraction of pole parameters it is essential that different reactions
are analyzed consistently. This requires especially that scattering as well as
different production reactions are studied together. However, the left hand cuts
of these different processes are different and as a result any factorization ansatz
that writes the production amplitude as being proportional to the scattering $T$--matrix,
as it is used, e.g., in $K$--matrix approaches,
has an intrinsic, uncontrollable uncertainty. A way around this is to either
keep all real parts in the loops, or to use dispersion theory. An example
of a consistent dispersive treatment of $\pi\pi$--scattering, and the pion
vector form factor that also allows for the explicit inclusion of higher resonances
is given in Ref.~\cite{piFF}.

Last but not least one should mention the role of chiral symmetry. As a consequence
of the Goldstone theorem pions should decouple from matter in the chiral limit
at vanishing momenta. As a consequence the interaction of pions with, e.g., resonances
should either go via derivatives or via quark mass insertions. In Ref.~\cite{achot} it
is shown that only then the resonances in the $S_{11}$ pion--nucleon channel decouple
from the threshold as required by chiral symmetry.

\newpage

\subsection{Study of baryon excited status at BES}
\addtocontents{toc}{\hspace{2cm}{\sl X.~Ji}\par}

\vspace{5mm}

\noindent
Xiaobin Ji

\vspace{5mm}

\noindent
Institute of High Energy Physics, CAS, Beijing, China\\

\vspace{5mm}
The report gives an experimental review of the baryon excited study at
BES. BES has its advantages to study baryon excited
states~\cite{Zou:2000re}. For $J/\psi\to N\bar N\pi$ and $J/\psi\to
N\bar N\pi\pi$, $N\pi$ and $N\pi\pi$ systems are limited to be pure
isospin 1/2 because the isospin conservation. And not only excited
baryon states, but also excited hyperons can be studied in the
charmonium decays. In the partial wave analysis, the data are fitted
applying an unbinned maximum likelihood fit. The amplitudes are
constructed using the relativistic covariant tensor amplitude
formalism~\cite{Rarita:1941mf, Liang:2002tk}. A powerful tool
FDC-PWA~\cite{Wang:2004du, FDC} had been developed to generate the
FORTRAN sources of PWA.

N* production in $J/\psi\to p\bar p\eta$ was studied with $7.8\times
10^6$ $J/\psi$ BESI events~\cite{Bai:2001ua}. Two N* resonances were
observed. In the analysis of $J/\psi\to p\bar n\pi^-$ + c.c., a
possible missing N* named N(2065) was observed with $5.8\times 10^7$
$J/\psi$ events at BESII~\cite{Ablikim:2004ug}. N(2065) was also
observed in the decay of $J/\psi\to p\bar p\pi^0$ with BESII data
samples~\cite{Ablikim:2009iw}. With $106\times 10^6$ $\psi(3686)$
events collected at BESIII~\cite{Ablikim:2009aa}, PWA of
$\psi(3686)\to p\bar p\pi^0$ was performed. Two new N* are observed,
N(2300)(1/2+) and N(2570)(5/2-)~\cite{Ablikim:2012zk}. PWA of
$\psi(3686)\to p\bar p\eta$ was performed either. Other baryon related
papers published by BESIII can be found in the
reference~\cite{Ablikim:2012ff, Ablikim:2012ih, Ablikim:2011uf}.

BESIII had collected more than 1.2 billion $J/\psi$, 0.5 billion
$\psi(3686)$ and 2.9 $fb^{-1}$ $\psi(3770)$, more and more results
can be expected.

\newpage

\subsection{Partial-Wave Analyses at Kent State University}
\addtocontents{toc}{\hspace{2cm}{\sl D.M.~Manley}\par}

\vspace{5mm}

\noindent
D. Mark Manley

\vspace{5mm}

\noindent
Department of Physics, Kent State University, Kent, OH, USA\\

\vspace{5mm}
An overview of the recent single-energy partial-wave analyses performed at
Kent State University is discussed.  These analyses included
data for $\overline{K}N$ scattering to $\overline{K}N$,
$\pi \Lambda$, and $\pi \Sigma$ final states
as well as data for $\pi N$ scattering to $\eta N$ and $K \Lambda$
final states.  Energy-dependent solutions were obtained by
incorporating our single-energy solutions into global multichannel
fits.  The resulting energy-dependent solutions provided a wealth
of information about $N^*$ and $\Delta$ resonances, as well as
information about $\Lambda^*$ and $\Sigma^*$ resonances.

Our $\overline{K}N$ analyses began with a multichannel fit~[1] of
published partial-wave amplitudes from scattering to various
final states, including $\overline{K}N$~[2], $\overline{K}^*(892)N$~[3],
$\overline{K}\Delta$~[4], $\pi \Lambda$~[2], $\pi \Lambda(1520)$~[5],
$\pi \Sigma$~[2], and $\pi\Sigma(1385)$~[6].  The channels $\sigma \Lambda$,
$\sigma \Sigma$, and $\eta \Sigma$ (for $S_{11}$) were included as
``dummy'' channels (channels without data) in order to satisfy unitarity
of the partial-wave $S$-matrix.  In addition, our fits included
Crystal Ball data~[7] for $K^- p \to \eta \Lambda$, which was used to determine the $S_{01}$ amplitude for the $\eta \Lambda$ channel. The resulting
global multichannel fits resulted in an energy-dependent solution that
described all of these various channels in a manner consistent
with $S$-matrix unitarity.

Our next step was to carry out single-energy partial-wave
analyses (PWAs).  Single-energy fits were performed separately for
(i) $K^- p \to K^- p$ and $K^- p \to \overline{K}^0 n$, for
(ii) $K^- p \to \pi^0 \Lambda$, and for
(iii) $K^- p \to \pi^+ \Sigma^-$, $K^- p \to \pi^0 \Sigma^0$,
and $K^- p \to \pi^- \Sigma^+$. Data were analyzed in c.m.\ energy
bins of width 20~MeV up to a maximum c.m.\ energy of 2100~MeV~[8].
Once we had a complete set of amplitudes from our own single-energy
PWAs, we carried out global multichannel energy-dependent fits using
a procedure similar to that in Ref.~[1].

More recently, single-energy partial-wave analyses up to a maximum
c.m.\ energy of about 2100~MeV have been completed for the
isospin-1/2 reactions
$\pi^- p \to \eta n$ and $\pi^- p \to K^0 \Lambda$~[9,10].  Data for these
reactions were analyzed in c.m.\ energy bins of width 30~MeV.
Global energy-dependent fits of the resulting amplitudes were performed using
a parametrization discussed in Ref.~[11].

The fits were constrained by including partial-wave amplitudes
for related reactions, including the SAID SP06 solution for
$\pi N \to \pi N$ [12], the SAID FA07 solution for $\gamma N \to \pi N$,
and the solution of Manley {\it et al.} [13] for $\pi N \to \pi \pi N$.
We found significant couplings of $S_{11}(1535)$ and $P_{11}(1710)$
to both $\eta N$ and $K \Lambda$.  Our results [14], in fact, confirmed the
existence of $P_{11}(1710)$ for which no evidence was found in the
analysis by Arndt {\it et al.} [12].  We found all
resonances, including a number of new states, recently
found independently in the Bonn-Gatchina analysis [15]. Our results
for $P_{11}(1880)$, $F_{15}(1860)$, $P_{13}(1900)$, and $D_{15}(2060)$
are in good agreement with those of the Bonn-Gatchina analysis, which
strengthens the evidence for these newly proposed states.

\newpage

\subsection{Hunting Resonance Poles with Rational Approximants}
\addtocontents{toc}{\hspace{2cm}{\sl P.~Masjuan}\par}

\vspace{5mm}

\noindent
P.~Masjuan

\vspace{5mm}

\noindent
Institut f\"ur Kernphysik, Johannes Gutenberg Universt\"at Mainz, Germany\\

\vspace{5mm}

The non-perturbative regime of QCD is characterized by the presence of physical resonances, complex poles of the amplitude in the transferred energy at higher (instead of the physical one) complex Riemann sheets. From the experimental point of view, one can obtain information about the spectral function of the amplitude through the time-like region ($q^2>0$) and also about its low-energy region through the experimental data on the space-like region ($q^2<0$).

Based on the mathematically well defined Pad\'{e} Theory \cite{Baker,PhDMasjuan}, we have developed a theoretically safe new procedure for the extraction of the pole mass and width of resonances~\cite{ProcJJ,Masjuan:2010ck,JJP}. The method uses a sequence of Pad\'e Approximants (PA) as a fitting functions to the available experimental data on the real axis to extract such parameters. PA are rational functions with coefficients identical to the Taylor expansion of the function to be approximated. Pad\'e Theory provides, then, with a set of convergence theorem for such sequence of approximants (for example, when the function obeys a dispersion relation given in terms of a positive definite spectral function, see~\cite{Baker,Peris:2006ds,MSCV,Masjuan:2009wy}; and when is a meromorphic function, see~\cite{pommerenke,Masjuan,Masjuan2} or \cite{Masjuan:2008xg} for a summary).

The standard procedure is to construct the PA from the Taylor expansion of the function at the origin of energies ($q^2=0$). When this expansion is not known, one can use a sequence of PA as a fitting functions to the experimental data to obtain such Taylor expansion up to a certain order  \cite{MJJP-VFF,Masjuan:2012wy}. Despite the nice convergence and the systematical treatment of the errors with this method, the procedure does not allow to obtain properties of the amplitude above the threshold. The reason is simple: the convergence of a sequence of PA centered at the origin of energies is limited by the presence of the production brunch cut~\cite{PhDMasjuan}. The PA sequence converge everywhere except on the cut, so one cannot access the Second Riemann sheet in such a way. Still, the mathematical Pad\'{e} Theory allows to produce a model-independent determination of resonance poles when certain conditions are fulfilled. The most important one is to center the PA sequence above the branch-cut singularity (beyond the first production threshold) instead of at origin of energies, and use time-like data for the fit. In this context one invokes the Montessus de Ballore's theorem \cite{montessus} to unfold the Second Riemann sheet of an amplitude and search the position of the resonance pole in the complex plane (for details see~\cite{ProcJJ,Masjuan:2010ck,JJP}). Our approach is similar to the one presented in \cite{a16} but with the expansion point at the real axis instead of around the pole position.

This model-independent method does not depend on a particular Lagrangian realization or modelization on how to extrapolate from the data on the real energy axis into the complex plane.

This work is performed in collaboration with Juan Jos\'{e} Sanz Cillero and supported by the Deutsche Forschungsgemeinschaft DFG through the Collaborative Research Center ÒThe Low-Energy Frontier of the Standard ModelÓ (SFB 1044).

\newpage

\subsection{A new method for extracting poles from single channel data - some results}
\addtocontents{toc}{\hspace{2cm}{ \sl H.~Osmanovic} \par }

\vspace{5mm}

\noindent
H. Osmanovic,$^1$ A. Svarc,$^2$ M. Hadzimehmedovic,$^1$ J. Stahov,$^1$ L. Tiator,$^3$ R. Workman$^4$

\vspace{5mm}

\noindent
$^1$ University of Tuzla, Faculty of Science, Bosnia and Herzegovina\\
$^2$ Rudjer Boskovic Institute, Zagreb, Croatia\\
$^3$Institut f\"ur Kernphysik, Johannes Gutenberg Universt\"at Mainz, Germany\\
$^4$ The George Washington University\\

\vspace{5mm}
We present results obtained from a new method [1] of extraction of poles of partial wave T- matrix.

As an input we  have used single energy (SE) and energy dependent (ED) solutions for electromagnetic multipoles from isobar model MAID [3]
and the model-independent GWU-SAID analysis [4].

There are many cuts and corresponding branchpoints in analytic structure of partial waves.

In the method, choosing of branchpoint positions is of crucial importance.
Our calculations show that a good and reliable fit to the partial wave data may be obtained by using three
Pietarinen expansions introducing three branchpoints and three corresponding cuts.

One effective cut and corresponding Pietarinen expansion is used to represent all cuts below physical $\pi N$ threshold.
Concerning remaining two Pietarinen expansions we applied two strategies.
In first strategy the branchpoints coincide with the physical thresholds ($\pi N$; $\pi \pi N$ or $\eta N$).
 This strategy is applied when fitting SE solutions.
 In the second strategy, branchpoint positions are free parameters in a fit and are effective branchpoints.
This strategy is applied when fitting ED solutions because, as a rule, such data assume certain, unknown analytic structure of partial wave.

The method gives reliable results for both input data sets (SE, ED solutions).

Presented  method is well suited for obtaining a global fit to the photoproduction and pion-nucleon scattering data,
and also  for extraction of baryon resonance parameters from photoproduction data and from single-channel pion-nucleon data as well.

\newpage

\subsection{From Raw Data to Resonances}
\addtocontents{toc}{\hspace{2cm}{\sl K.~Peters}\par}

\vspace{5mm}

\noindent
K. Peters

\vspace{5mm}

\noindent
GSI Darmstadt and Goethe University Frankfurt

\vspace{5mm}

Particle physics at small distances is well understood although this is not true for large distances when hadronization, light mesons and resonances are considered. These are barely understood compared to their abundance. A deeper understanding of the interaction and dynamics of light hadrons will improve the knowledge about non-perturbative QCD. We need appropriate parameterizations for the multi-particle phase space and a translation from parameters to effective degrees of freedom.

The analysis is usually performed by fitting models to the experimentally derived phase space distributions. Performing this kind of analysis is extremely difficult and error prone and a systematic treatment is both necessary and also very demanding. The goals are the unambiguous determination of masses, line-shapes and decay ratios and/or pole positions and couplings. In contrast to other complex optimization processes, like in technology issues and finance, it is important to target the correct solution and not just a good solution, since many solutions may end up with a fair numerical explanation of the data, while they differ dramatically in their physical content and interpretation.

The presentation covers a short review on the different steps in a concrete analysis process covering general considerations, the course of action in a realistic analysis and various detailed subtopics.

To arrive at the ultimate goal it has to be considered which processes take place (interactions, scales), what are the conserved properties (in term of kinematics and quantum numbers) and what are relevant parameters (order of magnitude and  relationships) and if factorization is a relevant aspect. The general course of actions  is very similar among the various analyses. It starts with data analysis, modeling of the interaction, the quality assured fitting to the data and finally the review and publication.

Important details are the consideration of relevant phase space observables and the treatment of various kinds of background. Since the necessary amplitudes are highly complex numerical and mathematical problems have to be investigated and solved for stable and verifiable solutions and unambiguous interpretation of the data.

Most aspects appear in almost every kind of reaction and therefore similar technologies are necessary and it is of great importance to form a unified community to exchange the experience in these analyses to shoulder the demand from the  emerging high statistics experiments all over the world.

\newpage

\subsection{The study of the proton-proton collisions at the beam momentum 2 GeV/c measured with HADES within Bonn-Gatchina PWA}
\addtocontents{toc}{\hspace{2cm}{\sl W.~Przygoda}\par}

\vspace{5mm}

\noindent
W. Przygoda for the HADES Collaboration

\vspace{5mm}

\noindent
Smoluchowski Institute of Physics, Jagiellonian University, 30-059 Cracow, Poland\\

\vspace{5mm}

HADES is a versatile magnetic spectrometer installed at GSI Darmstadt on SIS18 [1]. Thanks to its high acceptance, powerful particle ($p/K/\pi/e$) identification and very good mass resolution ($2-3\%$ for dielectrons in the light vector meson mass range) it allows to study both hadron and rare dilepton production in $N+N$, $p+A$,  $A+A$ collisions at a few AGeV beam energy range. In $p+p$ @ $1.25$ $GeV$, intermediate $\Delta(1232)$ resonance is expected to play a dominant role in the pion production but is not sufficient to describe fully the data. The resonance cross section was determined from exclusive $pp\pi^{0}$ and $np\pi^{+}$ channels [2] in the framework of OPE model with the accuracy of $20-30\%$. Investigation of these reaction channels by means of the PWA (Partial Wave Analysis) was also done [3] by the Bonn-Gatchina group at a smaller beam energy [4]. It revealed a dominant contribution of $\Delta(1232)p$ intermediate state but also sizeable non-resonant terms and interference effects.

In this work we report on the partial wave analysis of the single pion production in proton-proton collisions (as in [2]) measured with the HADES spectrometer. The $pp$ is pure isospin $I=1$ state and at this beam energy the following initial $pp$-states contribute (J=0) $^{1}S_{0}$, $^{3}P_{0}$, (J=1) $^{3}P_{1}$, (J=2) $^{1}D_{2}$, $^{3}P_{2}$, $^{3}F_{2}$. The higher total angular momentum contributions (i.e. $^{3}F_{3}$ for J=3) seem not to improve the obtained solutions. The final states are limited to $S-$, $P-$ and $D-$wave states with two possible intermediate resonance states $P_{33}(1232)$ and $P_{11}(1440)$. The data samples (for both channels $pp\pi^{0}$ and $pn\pi^{+}$) of 60000 events were analysed with the event-by-event background estimation (Q-factors). The analysis was preformed with other available data (see [5], 11 measurements for $pp\pi^{0}$ and only one for $pn\pi^{+}$ channel) covering mostly lower beam energies. The stability of solutions was investigated based on o few parametrisations of the transition amplitude $A_{tr}$ (with total energy dependence) and various descriptions of resonance states ($\Delta$ and $N^{*}$). The obtained solutions are still preliminary but generally describe the HADES data very well in various projection observables (CM angular distributions, invariant masses, angular distributions in the helicity and the Godfrey-Jackson frames). The analysis shows the dominant $P_{33}(1232)$ contribution in the $pp\pi^{0}$ at the level of $90\%$ ($\pm10\%$) which is an important message for the dilepton analysis where BR of $\Delta$ Dalitz decay can be identified (in the $pe^{+}e^{-}$ channel).

\newpage

\subsection{Positivity constraints in $\pi N$ scattering}
\addtocontents{toc}{\hspace{2cm}{\sl J.J. Sanz-Cillero}\par}

\vspace{5mm}

\noindent
J.J. Sanz-Cillero~\footnote{
I would like to thank the organizers for the nice scientific environment and
interesting discussions during the workshop.
In addition, I would like to thank the collaborators
D.L. Yao and H.Q. Zheng for their comments and suggestions
to the present note.}

\vspace{5mm}

\noindent
Departamento de F\'\i sica Te\'orica,
Universidad Aut\'onoma de Madrid, \\
Cantoblanco, 28049 Madrid, Spain\\

\vspace{5mm}

In the ongoing work presented here~\cite{ongoing} we propose the use of general S--matrix arguments such
as analiticity, crossing and unitarity, to constrain $\pi N$ interaction and its chiral
effective theory
description~\cite{Oller:2012}--\cite{op3-HBChPT-Fettes:1998}.
%%%
%%%description~\cite{Oller:2012,Yao:2012,op4-HBChPT-Fettes:2000,op4-HBChPT-Lag-Fettes:2000,op3-HBChPT-Fettes:1998}.
%%%
We follow a  procedure similar to that in $\pi\pi$--scattering. There one finds
that suitable combinations
of  isospin amplitudes become real and positive-definite within the ``upper part'' of the Mandelstam triangle
($0<t<4 M_\pi^2$, $s<4 M_\pi^2$, $u<4 M_\pi^2$)~\cite{Manohar:2008}--\cite{Distler:2007}.
%%%
%%%($0<t<4 M_\pi^2$, $s<4 M_\pi^2$, $u<4 M_\pi^2$)~\cite{Manohar:2008,Pennington:2008,Annanthanarayan:1995,Dita:1995,Distler:2007}.
%%%
This derivations are based on fixed--$t$ dispersion relations where the isospin combinations are arranged in
such a way that both   $s$ and crossed channels have positive-definite spectral functions.
This allows the extraction of bounds on particular combinations of
Chiral Perturbation Theory ($\chi$PT) couplings  and the check of the convergence of
the chiral expansion.

A previous study in the case of $\pi N$ forward scattering was done in Ref.~\cite{Luo:2007}.
Here we propose an improvement of the bounds through a full scanning of the
``upper part'' of the $\pi N$ Mandelstam triangle
($0<t<4 M_\pi^2$, $s<(m_N+M_\pi)^2$, $u<(m_N+M_\pi)^2$), not just the forward case $t=0$.
This should allow us to clarify if there is a clear improvement in the convergence
in Covariant Baryon $\chi$PT (CB$\chi$PT) when going from $\cO(p^3)$ to
$\cO(p^4)$~\cite{Oller:2012,Yao:2012}. Likewise, this can be relevant to constrain
the extension of CB$\chi$PT with the inclusion of the $\Delta$ resonance~\cite{Delta}
and to stabilize the phenomenological fits, providing an extra criterium to discern
``good'' and ``bad'' fit solutions.

The crucial ingredient for the positivity analysis is the partial wave (PW) decomposition of the
full scattering amplitude and its later reconstruction through  a summation of partial
waves~\cite{Ditsche:2012}. We write down a fixed--$t$ dispersion relation for the isospin amplitudes
and look for  combinations of the $\pi N$ scattering functions $A^I(\nu,t)$ and $B^I(\nu,t)$
(with $\nu\equiv (s-u)/(4 m_N)$)
that have a positive-definite $s$--channel spectral function. In a second step, one looks for
isospin combinations that arrange a positive definite $u$--channel spectral
function~\cite{Luo:2007,Meissner:2000}.

A first study of the $\cO(p^3)$ results
in pure CB$\chi$PT~\cite{Oller:2012,Yao:2012}, i.e., without the $\Delta$,
points out problems in the fulfillment of the extracted bounds for the
corresponding low-energy constants. At tree-level at this order one
can obtain the chiral coupling bound
\begin{eqnarray}
|6\nu d_3| &\leq c_2 \,\,\, +\,\,\, (\alpha-1)\, [
c_2\,  - \, 2m_N  (d_{14} -d_{15}]\, ,
\end{eqnarray}
with $|\alpha-1|\leq t/(4 m_N M_\pi)$ and $\nu$ and $t$ in the referred
``upper part''  of the Mandelstam triangle~\cite{ongoing}.
However, preliminary studies seem to hint an improvement
in the bound when the $\cO(p^3)$ one-loop contributions
are taken into account. This analysis is currently ongoing~\cite{ongoing}.
Furthermore, the $\cO(p^4)$ outcomes fulfill again the bounds at tree-level,
indicating how
the inclusion of higher chiral corrections improve the
convergence of the CB$\chi$PT expansion~\cite{Yao:2012}.
Likewise, an overview of previous Heavy Baryon $\chi$PT results
at $\cO(p^4)$~\cite{op4-HBChPT-Fettes:2000}       %%%,op4-HBChPT-Lag-Fettes:2000}
shows that the best fit solutions are those that
fulfill our positivity bounds.

\newpage

\subsection{Baryon spectroscopy with pion- and photon induced reactions}
\addtocontents{toc}{\hspace{2cm}{\sl V.~Shklyar}\par}

\vspace{5mm}

\noindent
V. Shklyar, H. Lenske, and U. Mosel

\vspace{5mm}

\noindent
Institut f\"ur Theoretische Physik I, Justus Liebig  Universt\"at Giessen, Germany\\

\vspace{5mm}
Lattice QCD calculations predict more states that observed from the analysis of
experimental data. This raises the question on the extraction of resonance
properties in scattering experiments. Since the most information on the baryon
spectra comes from the analysis of the pion-nucleon elastic scattering
it is necessary to investigated the less studied case of multiparticle production.
Thus the investigation e.g.  of the two-pion production would provide
an additional information on the excitation spectra  of the nucleon.

The aim of the present work is twofold.  First we extend the coupled-channel
unitary Lagrangian model (Giessen model) [1,2] to incorporate  two-pion final state
into analysis. The $2\pi N$ final state is  treated in the isobar
approximation with   $\sigma N $, $\pi \Delta $ , and , $\pi \Delta $ in the intermediate
subchannels. This formulation maintains two-body unitarity by construction.

Using the developed model we perform calculation of the $\pi^- p \to 2\pi^0 n$ production in the
first resonance energy region. One of the interesting questions here is the production
rate of the sigma meson due to the t-channel pion exchange.  Using the values
$m_\sigma =600\,$MeV and $\Gamma_\sigma=600$\,MeV the coupling constant $g_{\sigma \pi\pi}\approx 2$ has been
derived. The calculated $2\pi^0 N$ differential cross section was compared with the experimental
data [3]. Our results demonstrate a small contribution from the t-channel pion exchange to
the process  $\pi^- p \to \sigma N \to\pi^0\pi^0 n$ what  is in line  this conclusion made in [4].

A  good  description of the differential cross sections and  mass-distribution [4] can already be achieved
by including  the  $P_{11}(1440)$-resonance in to the calculations. We conclude that the contribution from
other states is small. However for c.m. energies above  1.5 GeV the effect from the   $D_{13}(1520)$ resonance
should also be taken into account.

\newpage

\subsection{Role of the Final State Interactions in Extraction of Interaction Parameters}
\addtocontents{toc}{\hspace{2cm}{\sl I.I.~Strakovsky}\par}

\vspace{5mm}

\noindent
I.I.~Strakovsky$^1$, W.J.~Briscoe$^1$, D.~Schott$^1$, R.L.~Workman$^1$, A.E.~Kudryavtsev$^{1,2}$, and V.E.~Tarasov$^2$

\vspace{5mm}

\noindent
$^1$The George Washington University, Washington, DC, USA\\
$^2$Institute of Theoretical and Experimental Physics,
Moscow, 117259 Russia\\

\vspace{5mm}
An accurate evaluation of the electromagnetic couplings
$N^\ast (\Delta^\ast)\to\gamma N$ from meson photoproduction
data remains a paramount task in hadron physics. A wealth of
new data for meson photoproduction is becoming available from
nuclear facilities worldwide. These measurements are now
beginning to have a significant impact on both the resonance
spectrum and its decay properties.

Here we focus on the single-pion production data and note
that a complete solution requires couplings from both charged
and neutral resonances, the latter requiring  $\pi^-p$ and
$\pi^0n$ photoproduction off a neutron target, typically a
neutron bound in a deuteron target. In addition to being less
precise, experimental data for neutron-target photoreactions
are much less abundant than those utilizing a proton target,
constituting only about 15\%
of the present World database~\cite{a2}. At low to
intermediate energies, this lack of neutron-target data is
partially compensated by experiments using pionic beams,
e.g., $\pi^-p\to\gamma n$, as has been measured.  However,
the disadvantage of using the reaction $\pi^-p\to\gamma n$
is the 5 to 500 times larger cross sections for $\pi^-p\to
\pi^0n\to\gamma\gamma n$, depending on photon energy,
$E_\gamma$, and pion production angle, $\theta$.

Extraction of the two-body ($\gamma n\to \pi^-p$ and $\gamma
n\to\pi^0 n$) cross sections requires the use of a
model-dependent nuclear correction, which mainly come from
final state interactions (FSI). As a result, our knowledge
of the neutral resonance couplings is less precise as
compared to the charged values~\cite{a1}.

We recently applied our FSI corrections~\cite{a3} to CLAS
$\gamma d\to\pi^-pp$ data~\cite{a4} to get elementary cross
sections for $\gamma p\to\pi^-p$ for the broad energy range,
$E_\gamma$=1000 through 2700~MeV~\cite{a5}.  Then we did the
same for the MAMI-B GDH experiment~\cite{a6} to get $\gamma
p\to\pi^-p$ around the $\Delta$-isobar ~\cite{a7}.

To accomplish a state-of-the-art analysis, we included FSI
corrections using a diagrammatic technique, taking into
account a kinematical cut with momenta less (more) than
200~MeV/$c$ for CLAS and 270~MeV/$c$ for MAMI-B for slow
(fast) outgoing protons.  The FSI correction factor for
both CLAS and MAMI-B kinematics was found to be small,
$\Delta\sigma/\sigma <$10\%.
However, these new cross sections departed significantly
from our predictions, at the higher energies, and greatly
modified the fit result.

New neutron A$_{1/2}$ and A$_{3/2}$ couplings shown
sometimes a significant deviation from our previous
determination and PDG average values, for instance, for
N(1650)$1/2^-$ and N(1675)$5/2^-$~\cite{a5}.

Connection between scattering and decay processes provides a
solid theoretical ground for describing some hadronic effects
$-$ ``Watson's theorem"~\cite{a8}.  For pion photoproduction,
isospin amplitudes have to satisfy Watson's theorem below the
2$\pi$-threshold allowing for a smooth departure from constraint
at high energies.  $$a\tan(\frac{ImA_l}{ReA_l})=
\delta_l(\pi N).$$  Above 2$\pi$-threshold, the rule may
still be true, if inelasticity of the corresponding
$\pi$N-elastic amplitude is small (as, e.g., for P$_{33}$.)

We evaluated results of several pion photoproduction analyses
such as SAID, MAID, EBAC, Giessen, and BnGa and compare $\pi$N
phases coming from $\pi$N and pion photoproduction amplitudes
on proton target. SAID uses $\pi N$ PWA results as a constraint
for analysis of pion photoproduction data.  Most of pion
photoproduction analyses are doing the same and uses SAID
$\pi$N outcome or its modification as input.

Our short summary is\\
%\begin{itemize}
i) Phases coming from $\pi$N amplitudes of different analyses
    are consistent.\\
ii) Some phases coming from different pion photoproduction
    analyses are consistent to each other and phases coming
    from $\pi$N amplitudes: \textit{3/2-Isospin Amplitudes}:
    E0+, E1+, and M1+. \\
iii) Some phases coming from different pion photoproduction
    analyses are inconsistent to each other and phases
    coming from $\pi$N amplitudes: \textit{3/2-Isospin
    Amplitudes}: M1-, E2-, M2-, E2+, and M2+; \textit{1/2-
    Isospin Amplitudes}: E0+, M1-, E1+, M1+, E2-, M2-, E2+,
    and M2+.\\
iv) Some phases coming from $E$ and $M$ multipoles are
    inconsistent  to each other and phases coming from
    $\pi$N amplitudes.\\
%\end{itemize}

%%%%%%%%%%%%%%%%%%%%%%%%%%%%%%%%%%%%%%%%%%%%%%%%%%%%%%%%%%%%%%%
\textbf{Acknowledgments}: This work was supported in part by
the U.S. DOE Grant No.~DE--FG02--99ER41110, by the Russian
RFBR Grant No.~02--0216465 and by the Russian Atomic Energy
Corporation ``Rosatom" Grant No.~NSb--4172.2010.2.

%%%%%%%%%%%%%%%%%%%%%%%%%%%%%%%%%%%%%%%%%%%%%%%%%%%%%%%%%%%

\newpage

\subsection{New single-channel pole extraction method}
\addtocontents{toc}{\hspace{2cm}{\sl  A.~\v{S}varc}\par}

\vspace{5mm}

\noindent
A. \v{S}varc$^1$, M. Had\v{z}imehmedovi\'{c}$^2$, H. Osmanovi\'{c}$^2$, J. Stahov$^2$

\vspace{5mm}
\noindent
$^1$ Rudjer Bo\v{s}kovi\'{c} Institute, Zagreb, Croatia\\
$^2$ University of Tuzla, Tuzla, Bosnia and Herzegovina

\vspace{5mm}
We present a new approach to quantifying pole parameters of single-channel processes based on Laurent expansion of partial wave T-matrices  %%@
\cite{a16}. Instead of guessing the analytical form of non-singular part of Laurent expansion as it is usually done, we represent it by the %%@
convergent series of Pietarinen functions with well defined analytic properties  \cite{Ciulli,Ciulli1,Pietarinen,Hoehler84}. As the analytic %%@
structure of non-singular term is usually very well known (physical cuts with branhcpoints at inelastic thresholds, and unphysical cuts in %%@
the negative energy plane), we show that we need one Pietarinen series per cut, and the number of terms in each Pietarinen series is %%@
automatically determined by the quality of the fit. The method is tested on a toy-model constructed from two known poles, various background %%@
terms, and two physical cuts, and shown to be robust and confident up to three Pietarinen series. We also apply this method to Zagreb CMB %%@
amplitudes for the N(1535) 1/2- resonance \cite{Batinic2010}, and confirm the applicability of the procedure for this case too.
Finally, we show preliminary results for fitting complete set of $\pi N$ elastic single energy partial wave amplitudes from GWU/SAID %%@
\cite{a2}, and a complete set of  single energy and energy dependent photoproduction multipoles from MAID \cite{a1} and GWU/SAID \cite{a2}.  %%@
The global agreement of the fit with all input data is always achieved, and the stability and reliability of extracted resonance parameters %%@
shows that this formalism can with full success be used for fitting a wide class of experimental data. The procedure is very similar as when %%@
Breit-Wigner functions are used, but with one modification: Laurent expansion with Pietarinen series is replacing the standard Breit-Wigner %%@
T-matrix form.

\newpage

\subsection{Hadron Spectroscopy: Supporting PWA at JLab }
\addtocontents{toc}{\hspace{2cm}{\sl  A.P.~Szczepaniak}\par}

\vspace{5mm}

\noindent
 A.P.~Szczepaniak
\vspace{5mm}

\noindent
Department of Physics, Indiana University, Bloomington IN 47405\\

\vspace{5mm}

I discuss goals of the new Physics Analysis Center that is being formed as a joint effort between JLab and Indiana University.
The center will oversee development of theoretical, phenomenological and computational tools for analysis of large data sets
from experiments in hadron spectroscopy. The center will facilitate interactions between practitioners in the physics of strong interactions by creating a forum for exchange of knowledge through collaboration meetings, workshops and summer schools.
The official operations of the Center will begin in the Fall of 2013.

One of the main physics goals in strong interactions physics is to identify gluonic excitations, and specifically,  manifestations of
 gluonic degrees of freedom in the spectrum of hadrons~\cite{Crede:2008vw,Klempt:2007cp,Meyer:2010ku}.  Lattice gauge simulations\cite{Dudek:2009qf,Dudek:2011bn,Dudek:2012vr}  have now produced the spectrum of gluonic excitations which appear to have multiplet structure analogous to that of the quark model for ordinary hadrons \cite{Guo:2008yz,Szczepaniak:2006nx}.
There is strong experimental evidence, in at least one case; the $\eta'\pi$ production of exotic partial wave with resonant characteristics~\cite{Beladidze:1993km, Ivanov:2001rv,Schluter:2012re}

The immediate goal of the Physics Analysis Center will be to provide theoretical support needed to construct reaction amplitudes that fulfill the S-matrix matrix constraints, and take advantage of modern development in lattice gauge simulations and other theoretical approaches~\cite{Dudek:2011tt}.

\newpage

\subsection{Complete Experiments for PWA and Baryon Resonance Extraction}
\addtocontents{toc}{\hspace{2cm}{\sl L.~Tiator}\par}

\vspace{5mm}

\noindent
L.~Tiator

\vspace{5mm}

\noindent
Institut f\"ur Kernphysik, Johannes Gutenberg Universt\"at Mainz, Germany\\

\vspace{5mm}

The current status of partial wave analysis, mainly from single-pion
photoproduction on protons, is discussed and results of the isobar
model MAID~[1], the model-independent GWU-SAID~[2] analysis and the
Bonn-Gatchina~[3] coupled-channels analysis are compared. Except for
a few dominant partial waves, as P33, most other waves disagree
among the different analyses, which can mainly attributed to the
incomplete data base of photoproduction observables.

To improve this situation, possibilities of a model-independent
partial wave analysis for pion, eta or kaon photoproduction are
discussed in the context of `complete experiments'. It is shown that
the helicity amplitudes obtained from at least 8 polarization
observables~[4-7] including beam, target and recoil polarization can
not be used to analyze nucleon resonances. However, a truncated
partial wave analysis, which requires only 5 observables will be
possible with minimal model assumptions~[8-12].

The result of such complete experiments will be almost model
independent single-energy partial wave amplitudes, which will be the
ideal starting point for baryon resonance extraction by using models
like isobar or dynamical models in single-channel or coupled channel
analysis. Furthermore, pole extraction methods, which are discussed
during this workshop can be applied as speed-plot~[14] and
time-delay techniques, regularization method~[15], Laurent
expansion, Pietarinen expansion~[16], Pade approximation~[17], pole
fitting methods~[18] and, whenever possible, analytical continuation
into the complex region.

\newpage

\subsection{Studies on a complete experiment for pseudoscalar meson photoproduction}
\addtocontents{toc}{\hspace{2cm}{\sl Y.~Wunderlich}\par}

\vspace{5mm}

\noindent
Y. Wunderlich

\vspace{5mm}

\noindent
Helmholtz-Institut f\"ur Strahlen- und Kernphysik, Universit\"at Bonn, Germany\\

\vspace{5mm}

In the reaction of photoproduction of single pseudoscalar mesons, there exists the possibility of measuring 16 distinct polarization observables for every fixed value of both energy and angle $(W, \hspace*{1pt} \theta)$ [1]. One observable is the unpolarized differential cross section, however there also exist 3 single polarization observables and 12 double polarization observables grouped into the classes beam-target, beam-recoil and target-recoil. It is anticipated that the in principle accessible volume of experimental information will aid the disentanglement of the strongly overlapping resonances of the nucleon.\\
Since the beginning of the 1970s the problem of the so called complete experiment started to emerge in the literature [2]. It consists of the question which minimum subsets of all 16 polarization observables are sufficient in order to maximally constrain the underlying amplitudes. This optimization problem is important in the context of currently ongoing polarization measurements at facilities like ELSA, MAMI and JLAB.\\
In the 1990s it was shown that 8 carefully chosen observables suffice to yield a complete experiment [3]. However, in the low energy area of the reaction and in connection to a maximally model independent truncated partial wave analysis, there exists the realistic chance for achieving completeness with even less than 8 observables [4]. \\
The above mentioned results are discussed and a new preliminary study on the ambiguities of the single spin observables, which was performed using the formalism of reference [4], is presented.

\newpage

\newpage

\section{List of participants}

\begin{flushleft}
\begin{itemize}
\item H.J.~Arends, UniversitŠt Mainz {\tt arends@kph.uni-mainz.de}
\item R.~Beck, UniversitŠt Bonn {\tt beck@hiskp.uni-bonn.de }
\item K.~Bicker, CERN {\tt kbicker@cern.ch }
\item J.~Biernat, University of Krak—w {\tt jacek.b.biernat@gmail.com }
\item W.~Briscoe, The George Washington University {\tt briscoe@gwu.edu }
\item P.~Buehler, Stefan Meyer Institut Wien {\tt paul.buehler@oeaw.ac.at }
\item X.~Cao, Universit\"at Giessen {\tt Xu.Cao@theo.physik.uni-giessen.de }
\item S.~Ceci, Rudjer Boskovic Institute Zagreb {\tt sasa.ceci@irb.hr}
\item A.~Denig, Universit\"at Mainz {\tt denig@kph.uni-mainz.de}
\item M.~D\"oring, Universit\"at Bonn {\tt doering@hiskp.uni-bonn.de}
\item E.~Downie, The George Washington University {\tt evie.downie@gmail.com}
\item E.~Epple, Technische Universit\"at M\"unchen {\tt eliane.epple@ph.tum.de }
\item C. Fern\'andez-Ram\'{\i}rez, Universidad Complutense de Madrid {\tt cefera@gmail.com}
\item A.~Fix, Tomsk Polytechnic University {\tt fix@tpu.ru}
\item M.~Fritsch, Universit\"at Mainz {\tt fritsch@kph.uni-mainz.de }
\item J.~Gegelia, Universit\"at Bochum {\tt Jambul.Gegelia@tp2.ruhr-uni-bochum.de }
\item M.~Gorshteyn, Universit\"at Mainz {\tt gorshtey@kph.uni-mainz.de }
\item A.~Gillitzer, Forschungszentrum JŸlich {\tt a.gillitzer@fz-juelich.de}
\item R.~Gothe, University of South Carolina {\tt gothe@sc.edu }
\item W.~Gradl, Universit\"at Mainz {\tt gradl@kph.uni-mainz.de }
\item B.~Grube, Technische Universit\"at M\"unchen {\tt bgrube@tum.de}
\item H.~Haberzettl, The George Washington University {\tt helmut@gwu.edu }
\item C.~Hanhart, Forschungszentrum JŸlich {\tt c.hanhart@fz-juelich.de}
\item A.~Hayrapetyan, Universit\"at Giessen {\tt Avetik.Hayrapetyan@uni-giessen.de }
\item P.~Jasinski, Helmholtz Institut Mainz {\tt jasinski@kph.uni-mainz.de }
\item X.~Ji, IHEP Beijing {\tt jixb@ihep.ac.cn }
\item E.M.~Kabuss, Universit\"at Mainz {\tt emk@kpk.uni-mainz.de }
\item A.~Karavdina, Universit\"at Mainz {\tt karavdin@kph.uni-mainz.de }
\item Y.~Liang, Universit\"at Giessen {\tt yutie.liang@physik.uni-giessen.de }
\item B.~Liu, IHEP Beijing {\tt liubj@ihep.ac.cn }
\item M.~Lutz, GSI Darmstadt {\tt m.lutz@gsi.de }
\item D.M.~Manley, Kent State University {\tt manley@kent.edu}
\item P.~Masjuan, Universt\"at Mainz, {\tt masjuan@kph.uni-mainz.de}
\item F.~Nerling, Helmholtz Institut Mainz {\tt f.nerling@gsi.de }
\item H.~Merkel, Universit\"at Mainz {\tt merkel@kph.uni-mainz.de }
\item H.~Osmanovic, University of Tuzla {\tt hedim.osmanovic@untz.ba}
\item K.~Peters, GSI Darmstadt {\tt k.peters@gsi.de }
\item A.~Pitka, Universit\"at Giessen {\tt pitka@physik.uni-giessen.de }
\item W.~Przygoda, University of Krakow {\tt przygoda@if.uj.edu.pl}
\item A.~Sanchez-Lorente, Helmholtz Institut Mainz {\tt a.sanchez@gsi.de }
\item J.J.~Sanz-Cillero, Universidad Aut\'onoma de Madrid {\tt cillero@ifae.es}
\item A.~Sarantsev, Universit\"at Bonn {\tt andsar@hiskp.uni-bonn.de }
\item S.~Scherer, Universit\"at Mainz {\tt scherer@kph.uni-mainz.de }
\item K.~Schoenning, University of Uppsala {\tt karin.schonning@cern.ch }
\item D.~Schott, The George Washington University {\tt dschott@email.gwu.edu }
\item V.~Shklyar, Universit\"at Giessen {\tt shklyar@theo.physik.uni-giessen.de }
\item I.I.~Strakovsky, The George Washington University {\tt igor@gwu.edu}
\item A. \v{S}varc, University of Tuzla {\tt svarc@irb.hr}
\item A.~Szczepaniak, University of Indiana {\tt aszczepa@indiana.edu }
\item U.~Thoma, Universit\"at Bonn {\tt thoma@hiskp.uni-bonn.de }
\item A.~Thomas, Universit\"at Mainz {\tt thomas@kph.uni-mainz.de }
\item L.~Tiator, Universt\"at Mainz {\tt tiator@kph.uni-mainz.de}
\item M.~Vanderhaeghen, Universit\"at Mainz {\tt marcvdh@kph.uni-mainz.de }
\item Weisrock, Universit\"at Mainz {\tt weisrock@kph.uni-mainz.de }
\item P.~Weidenkaff, Universit\"at Mainz {\tt weidenka@kph.uni-mainz.de }
\item Y.~Wunderlich, Universit\"at Bonn {\tt wunderlich@hiskp.uni-bonn.de }
\item H.~Ye, Universit\"at Giessen {\tt yehua@ihep.ac.cn }
\item Q.~Zhao, IHEP Beijing {\tt zhaoq@ihep.ac.cn }
\end{itemize}
\end{flushleft}

\end{document}